\documentclass{iopart}
\usepackage{amssymb,mathptmx,times,rangcite,graphicx}
\usepackage[varg]{txfonts}
\long\def\@makefntext#1{\parindent 1em\noindent
 \makebox[1em][l]{\footnotesize\rm$^\arabic{footnote}$}%
 \footnotesize\rm #1}
\def\@makefnmark{\hbox{$^\arabic{footnote}$}}
\def\@thefnmark{\arabic{footnote}}

\def\be{\begin{equation}}
\def\ee{\end{equation}}
\def\bea{\begin{eqnarray}}
\def\eea{\end{eqnarray}}
\def\bse{\numparts}
\def\ese{\endnumparts}
\eqnobysec
\def\numparts{\refstepcounter{equation}%
     \setcounter{eqnval}{\value{equation}}%
     \setcounter{equation}{0}%
     \def\theequation{\arabic{section}.\arabic{eqnval}{\it\alph{equation}}}}
\def\endnumparts{\def\theequation{\arabic{section}.\arabic{equation}}%
     \setcounter{equation}{\value{eqnval}}}

\let\truesum=\sum
\let\trueint=\int
\let\trueoint=\oint
\let\trueprod=\prod
\let\truecirc=\circ
\def\sum{\mathop{\textstyle\truesum}\limits}
\def\int{\mathop{\textstyle\trueint}\limits}
\def\oint{\mathop{\textstyle\trueoint}\limits}
\def\prod{\mathop{\textstyle\trueprod}\limits}
\def\circ{{\ifmmode\truecirc\else$\truecirc$\fi}}

\def\overl@ss#1#2{\vcenter{\offinterlineskip
        \ialign{$\m@th#1\hfil##\hfil$\crcr#2\crcr<\crcr } }}
\def\overgr@at#1#2{\vcenter{\offinterlineskip
        \ialign{$\m@th#1\hfil##\hfil$\crcr#2\crcr>\crcr } }}
\def\gl{\mathrel{\mathpalette\overl@ss>}}
\def\lg{\mathrel{\mathpalette\overgr@at<}}
\def\fbf#1{\setbox0=\hbox{$#1$}\kern-0.10\wd0
  \lower0.04em\copy0\kern-\wd0 \lower0.04em\hbox{\kern+0.05em\copy0}\kern-\wd0
  \raise0.00em\copy0\kern-\wd0 \raise0.00em\hbox{\kern-0.05em\box0}}
\renewenvironment{pmatrix}{\left(\!\!\begin{array}{cc}}{\end{array}\!\!\right)}
\def\_#1{{\mathsf{#1}}}
\def\@#1{{\mathbf{#1}}}
\makeatother
\def\Real{{\mathbb{R}}}

\def\Re{\mathop{\rm Re}\nolimits}
\def\Im{\mathop{\rm Im}\nolimits}

\def\diag{\mathop{\rm diag}\nolimits}
\def\Res{\mathop{\rm Res}\limits}
\def\sech{\mathop{\rm sech}\nolimits}
\let\.=\dot
\let\^=\hat
\let\~=\tilde
\let\==\bar
\let\#=\rm
\def\e{{\rm e}}
\def\d{{\rm d}}
\def\o#1{^{(#1)}}
\let\epsilon=\varepsilon

\def\punct{^{\,[\raise0.08ex\hbox{\scriptsize$\slash$}\kern-0.34em0]}}
\def\ext{{\mathrm{ext}}}
\advance\parskip 1pt

\begin{document}
\title[Solitons, BVPs and a nonlinear method of images]%
{Solitons, boundary value problems and a nonlinear\\method of images}
\author{Gino Biondini and Guenbo Hwang}
\address{State University of New York at Buffalo, Department of Mathematics, Buffalo, NY 14260}
\date{\small\today}

\begin{abstract}

\noindent 
We characterize the soliton solutions of the nonlinear Schr\"odinger 
equation on the half line with linearizable boundary conditions.
Using an extension of the solution to the whole line and the 
corresponding symmetries of the scattering data, 
we identify the properties of the discrete spectrum of the 
scattering problem.
We show that discrete eigenvalues appear in quartets as opposed to
pairs in the initial value problem, and we obtain explicit relations 
for the norming constants associated to symmetric eigenvalues.
The apparent reflection of each soliton at the boundary of the 
spatial domain is due to the presence of a ``mirror'' soliton, 
with equal amplitude and opposite velocity, located beyond the boundary.
We then calculate the position shift of the physical solitons as a 
result of the nonlinear reflection.
These results provide a nonlinear analogue of the method of images 
that is used to solve boundary value problems in electrostatics.
\par\medskip\noindent\today
\end{abstract}

\section{Introduction}

One of the hallmarks of integrability of a nonlinear evolution equation (NLEE)
is the existence of exact $N$-soliton solutions.
It is well known that
each soliton is associated to a discrete eigenvalue for 
the scattering problem for the given NLEE
via the inverse scattering transform (IST).
This has long been known to be true for initial value problems (IVPs) 
posed on an infinitely extended spatial domain
(e.g., see \cite{AblowitzClarkson,AblowitzSegur}).
Recent developments on the IST for initial-boundary value problems (IBVPs),
however,
have shown that the same statement also applies for problems posed over
a semi-infinite interval
\cite{JMP16p1054,IP24p065011,IP16p1813,JETPL74p481,PhysD35p167,JPA34p7539,JMP41p414}. 
The purpose of this work is to characterize the soliton solutions of IBVPs
for integrable NLEEs.

For concreteness, we consider the nonlinear Schr\"odinger (NLS) equation 
 \be
 iq_t+q_{xx}-2\nu|q|^2q=0\,,
 \label{e:NLS}
 \ee
where as usual $\nu=-1$ and $\nu=1$ denote respectively the 
focusing and defocusing cases. 
The IST for~\eref{e:NLS} on $-\infty<x<\infty$ was formulated 
in \cite{JETP34p62},
where the soliton solutions in the focusing case were also obtained,
including the well-known one-soliton solution
 \be \label{e:ivpsoliton}
 q_{\rm s}(x,t)= A\e^{i[Vx + (A^2-V^2)t +\varphi]} \sech[A(x-2Vt- \xi)]\,,
 \ee
where $k= (V + i\,A)/2$ is the discrete eigenvalue.
The IBVP for~\eref{e:NLS} on $0<x<\infty$
with homogeneous Dirichlet or Neumann boundary conditions (BCs)
at the origin was studied in \cite{JMP16p1054}
using the IST on the whole line and an odd or even extension of 
the potential, respectively.
The case of homogeneous Robin BCs,
 \be
 q_x(0,t)-\alpha q(0,t)=0\,,
 \label{e:RobinBCs}
 \ee
with $\alpha\in\Real$, was also studied in \cite{JPA24p2507,PhysD35p167,IP7p435}
using a clever extension of the potential to the whole line.
Recently, a new spectral method was proposed for the solution of IBVPs
for integrable NLEEs
\cite{PRSLA453p1411,JMP41p4188,NLTY18p1771}.
The method applies for generic BCs.  
In general, the solution of the IBVP 
requires solving a system of coupled nonlinear ordinary differential equations 
involving the spectral parameter in order to eliminate 
the unknown boundary data.
The method, however, also identifies a class of \textit{linearizable} BCs.
These linearizable BCs, which for the NLS equation
coincide with \eref{e:RobinBCs},
allow one to completely linearize the problem and to express the
solution of the IBVP as effectively as for the IVP.

Importantly, in all of the above methods the relation between 
solitons and discrete eigenvalues that exists in the IVP is preserved
in the IBVP,
yielding solutions of the form \eref{e:ivpsoliton}.
This leads to an apparent paradox, however, since 
\eref{e:ivpsoliton} does not satisfy the BCs~\eref{e:RobinBCs}. 
A further paradox is that
numerical solutions of the IBVP for~\eref{e:NLS}
show inequivocally that solitons are reflected at the boundary.
But the soliton velocity is the real part of the discrete eigenvalue, 
which does not change in time.
As we show below, the resolution of these paradoxes is that
discrete eigenvalues in the IBVP appear in \textit{quartets}, as opposed to
pairs in the IVP.
This means that, for each soliton in the physical domain (in our case,
the positive $x$-axis), a symmetric counterpart exists (i.e., on
the negative $x$-axis), with equal amplitude and opposite velocity, 
whose presence ensures that the whole solution satisfies the BCs.
The ostensible reflection of the soliton at the boundary of the physical
domain (here $x=0$) then corresponds simply to the interchanging of
roles between the ``true'' and ``mirror'' solitons.

It is worth noting that the method to obtain soliton solutions for the IBVP
on the half line is similar in spirit to the method of images that is
used to solve boundary value problems in electrostatics~\cite{Jackson}. 
Here, however, unlike the case of electrostatics,
the reflection experienced by the solitons comes accompanied by a 
corresponding position shift, which is a reminder of the nonlinear 
nature of the problem.

The outline of this work is the following. 
In section~\ref{s:IBVPNLS}
we discuss the IST for \eref{e:NLS} on the half line 
with linearizable BCs,
and we derive the symmetries of the discrete eigenvalues for the
scattering problem. 
In section~\ref{s:normingconst} we obtain the precise relations between 
discrete eigenvalues and norming constants.
In section~\ref{s:solitons} we discuss the behavior of the solitons
and we compute the shift originating from the reflection at the boundary, 
showing that this shift depends on the BCs.
Finally, section~\ref{s:discussion} concludes with some final remarks.


\section{Soliton solutions of the NLS equation on the half line}
\label{s:IBVPNLS}

Consider the IBVP for the NLS equation on the half line
with linearizable BCs; 
that is, \eref{e:NLS} on $0< x < \infty$ and $t>0$ 
and with \eref{e:RobinBCs} given.
When $\alpha =0$ or $\alpha\to \infty$, the BCs reduce to the
Dirichlet or the Neumann BCs: 
$q(0,t)=0$ and $q_x(0,t)=0$, respectively.
It is well-known that 
the NLS equation is the compatibility condition of the matrix Lax pair 
\cite{JETP34p62,PhysD35p167}
 \be \label{e:NLSLP}
 \mu_x-ik[\sigma_3,\mu] = \_Q\mu \,,\qquad
 \mu_t+2ik^2[\sigma_3,\mu] = \_H\mu \,,
 \ee
where $[\_A,\_B]=\_A\_B -\_B\_A$ is the matrix commutator, and
 \bse 
 \bea 
 \_Q(x,t)=
   \begin{pmatrix}0 &q(x,t)\\ r(x,t)&0\end{pmatrix}, \qquad 
 \sigma_3=
   \begin{pmatrix}1 &0\\0 &-1\end{pmatrix}\,,\\
 \_H(x,t,k)= -i\_Q\_Q\,\sigma_3-i\_Q_x\sigma_3-2k\_Q=
   \begin{pmatrix} -iqr & iq_x-2kq \\ -ir_x-2kr & iqr \end{pmatrix}\,,
 \eea 
 \ese
with $r(x,t)= \nu q^*(x,t)$ and where the asterisk denotes complex conjugation.
As usual, 
we assume that $q(x,0)$ is sufficiently regular and decays sufficiently fast
as $x\to\infty$ that Jost solutions and other relevant quantities are 
well-defined.

\subsection{IST for the NLS equation on the whole line}
\label{s:ISTIVP}

In sections~\ref{s:DirNeuBCs} and~\ref{s:RobinBCs} 
we characterize the solution of the IBVP for the NLS equation
on the half line using the IST for the problem on the whole line 
and an appropriate extension of the potential. 
Here we therefore briefly introduce the relevant quantities 
that will be used later.
We refer the reader to \cite{APT2003,AblowitzSegur,IP24p065011,JETP34p62} 
for all details.

We define the Jost solutions of~\eref{e:NLSLP} as the simultaneous solutions 
of both parts of the Lax pair~\eref{e:NLSLP}
that reduce to the identity matrix as $x\to \mp\infty$: 
for all $k\in\Real$,
 \bse \label{e:JostNLS} \bea
 \mu\o1(x,t,k)=
 \_I + \int_{-\infty}^x \e^{ik(x-x')\sigma_3}
   \_Q(x',t)\mu\o1(x',t,k)\e^{-ik(x-x')\sigma_3}\,\d x'\,,\\
 \mu\o2(x,t,k)=\_I - \int_x^\infty \e^{ik(x-x')\sigma_3}
   \_Q(x',t)\mu\o2(x',t,k)\e^{-ik(x-x')\sigma_3}\,\d x'\,.
 \eea \ese
We denote the columns of the eigenfunctions as 
$\mu\o{m}(x,t,k) =\big(\mu\o{m,L},\mu\o{m,R}\big)$, $m=1,2$.
The regions of analyticity of the Jost solutions, as effected by 
the exponentials in~\eref{e:JostNLS}, are \cite{APT2003,JETP34p62}:
 \be
  \mu\o{1,L},~\mu\o{2,R}\!:\quad \Im k<0\,,\qquad 
  \mu\o{1,R},~\mu\o{2,L}\!:\quad \Im k>0\,.
 \ee
Since $\det\mu\o{m}(x,t,k)=1$ for $m=1,2$, 
for all $k\in\Real$ $\mu\o1(x,t,k)$ and $\mu\o2(x,t,k)$ 
are both fundamental solutions of~\eref{e:NLSLP}. 
Hence, for all $k\in\Real$, we have the scattering relation:
 \be 
 \mu\o1(x,t,k)=
 \mu\o2(x,t,k)\,\e^{i\theta\sigma_3}\_A(k)\e^{-i\theta\sigma_3}\,,
 \label{e:scatteringNLS}
 \ee
where $\theta(x,t,k)= kx - 2k^2t$.
The limit of \eref{e:scatteringNLS} as $x\to\infty$ yields,
using \eref{e:JostNLS}, an integral representation for the
scattering matrix $\_A(k)$:
 \be 
 \_A(k)= \_I + \int_{-\infty}^\infty
 \e^{-i(kx-2k^2t)\sigma_3}\_Q(x,t)\mu\o1(x,t,k)\e^{i(kx-2k^2t)\sigma_3}\,\d x\,,
 \label{e:NLFT}
 \ee
In turn, \eref{e:NLFT} can be used to establish that 
the elements $a_{11}(k)$ and $a_{22}(k)$ of $\_A(k)$ 
can be analytically continued on $\Im k<0$ and $\Im k>0$, 
respectively.
Note that with the above definitions, 
$\_A(k)$ is independent of time.

The eigenfunctions and scattering coefficients obey the following 
symmetry relations: 
 \bse 
 \label{e:NLSsymmmu}
 \bea
 \mu\o{m,L}(x,t,k)=\sigma_\nu\mu\o{m,R}(x,t,k^*)^*\!,\quad
 \mu\o{m,R}(x,t,k)=\nu\sigma_\nu\mu\o{m,L}(x,t,k^*)^*\!,
 \\[-1ex]
\noalign{\noindent for $m=1,2$, where}
 \sigma_\nu= \begin{pmatrix} 0 & 1\\ \nu & 0 \end{pmatrix}\,,
 \nonumber\\
\noalign{\noindent together with}
 a_{22}(k)=a_{11}^*(k^*)\,,\qquad a_{21}(k)=\nu a_{12}^*(k)\,,
 \label{e:NLSsymma}
 \eea 
 \ese
Equations~\eref{e:NLSsymmmu} 
hold for all values of $k$ for which all terms are well-defined.
As a result, one can write the coefficients of $\_A(k)$ as
 \be 
 \_A(k)=\begin{pmatrix} a^*(k^*) & b(k) \\ \nu b^*(k^*)  &a(k)
 \end{pmatrix}\,.
 \ee

Discrete eigenvalues occur when $a_{11}(k)=0$ or $a_{22}(k)=0$~\cite{APT2003}. 
Since in the defocusing case $\nu=1$ with vanishing 
BCs at infinity there are no discrete eigenvalues \cite{JETP34p62}, 
whenever we discuss the discrete spectrum we implicitly set $\nu=-1$. 
Assuming that $a_{11}(k)a_{22}(k)\ne0$ $\forall k\in\Real$, 
there exist a finite number of such zeros,
since the scattering coefficients are sectionally analytic. 
We also assume such zeros are simple. 
Let us denote $k_j$ for $j=1,\dots,J$ and $\=k_j$ for $j=1,\dots,\=J$
the zeros of $a_{22}(k)$ and $a_{11}(k)$, respectively, 
where $\Im k_j > 0$ and $\Im \=k_j < 0$.
\unskip\break
The asymptotic behavior of the Jost solutions implies that
discrete eigenvalues occur when the decaying eigenfunctions
as $x\to-\infty$ are proportional to those as $x\to\infty$.  
That is,
 \label{e:eigenval}
 \bea \fl
 \mu\o{1,R}(x,t,k_j)=b_j\e^{2i\theta(x,t,k_j)}\mu\o{2,L}(x,t,k_j)\,,\quad
 \mu\o{1,L}(x,t,\=k_j)=\=b_j\e^{-2i\theta(x,t,\=k_j)}\mu\o{2,R}(x,t,\=k_j)\,.
 \eea
One then obtains the following residue relations:
 \bse
 \label{e:NLSnorming}
 \bea 
\fl
 \Res_{k=k_j}\bigg[\frac{\mu\o{1,R}}{a_{22}}\bigg]
 =C_j\e^{2i\theta(x,t,k_j)}\mu\o{2,L}(x,t,k_j)\,,\quad
 \Res_{k=\=k_j}\bigg[\frac{\mu\o{1,L}}{a_{11}}\bigg]=
 \=C_j\e^{-2i\theta(x,t,\=k_j)}\mu\o{2,R}(x,t,\=k_j)\,,
 \eea
where
 \be
 C_j=b_j/\.a_{22}(k_j)\,,\qquad
 \=C_j=\=b_j/\.a_{11}(\=k_j)\,,
 \ee
 \ese
and the overdot denotes differentiation.
As usual, $C_j$ and $\=C_j$ (or equivalently $b_j$ and $\=b_j$)
are referred to as the norming constants. 
The symmetry relations \eref{e:NLSsymmmu} imply
 \be
 \=J=J\,,\qquad 
 \=k_j=k_j^*\,,\qquad
 \=b_j=- b_j^*\,,\qquad
 \=C_j=-C_j^*\,.
 \label{e:NLSsymm0}
 \ee

To recover the potential from the scattering data one uses~\eref{e:scatteringNLS}
to define the matrix Riemann-Hilbert problem (RHP)
 \be 
 \_M^+(x,t,k)-\_M^-(x,t,k)= \_M^+(x,t,k)\_R(x,t,k)\,,
 \label{e:NLSRHP}
 \ee
for all $k\in\Real$,
where the matrix-valued sectionally meromorphic functions 
$\_M^\pm(x,t,k)$ are
 \bea \fl
 \_M^+(x,t,k)=
 \bigg(\mu\o{2,L}(x,t,k)\,,\frac{\mu\o{1,R}(x,t,k)}{a(k)}\bigg)\,,
 \quad 
 \_M^-(x,t,k)=
 \bigg(\frac{\mu\o{1,L}(x,t,k)}{a^*(k^*)}\,,\,\mu\o{2,R}(x,t,k)\bigg)\,,
\label{e:NLSMpmdef}
 \\
\noalign{\noindent the jump matrix is}
 \_R(x,t,k)=
 \begin{pmatrix}\nu|\rho(k)|^2& \e^{2i\theta(x,t,k)}\rho(k)\\
 -\nu\e^{-2i\theta(x,t,k)}\rho^*(k)& 0\end{pmatrix},
 \eea
and the reflection coefficient is
$\rho(k)= {b(k)}/{a(k)}$\,
for all $k\in\Real$.
Since $\mu\o{m}(x,t,k)=\_I+O(1/k)$ as $k\to\infty$ for $m=1,2$~\cite{APT2003},
the RHP~\eref{e:NLSRHP} is solved using standard Cauchy projectors,
after regularizing by subtracting the pole contributions from the 
discrete spectrum:
 \be \fl
 \_M(x,t,k)= \_I + \sum_{j=1}^J \bigg(
    \frac1{k-k_j} \Res_{k=k_j}\big[\_M^+\big]
    + \frac1{k- k_j^*}\Res_{k= k_j^*}\big[\_M^-\big] \bigg)
  + \frac1{2\pi i}\int_{-\infty}^\infty
    \_M^+(x,t,k')\frac{\_R(x,t,k')}{k'-k}\d k'. 
 \label{e:NLSRHPsoln}
 \ee
The asymptotic behavior of $\_M^\pm(x,t,k)$ then yields the 
reconstruction formula for the potential as
$q(x,t)=-2i\lim_{k\to \infty} k\,\_M_{12}^{\pm}(x,t,k)$:
 \bea\fl
 q(x,t)= -2i\sum_{j=1}^J
 C_j\,\e^{2i\theta(x,t,k_j)}\mu\o2_{11}(x,t,k_j)+
 \frac1\pi \int_{-\infty}^\infty \e^{2i\theta(x,t,k)}\rho(k)
 \mu\o2_{11}(x,t,k)\,\d k\,.
 \label{e:reconstq}
 \eea
Hereafter we will often write the norming constants as 
$C_j= A_j\e^{A_j\xi_j+i(\varphi_j+\pi/2)}$ for $j=1,\dots,J$.
 
In the reflectionless case [i.e., $\rho(k)=0$ $\forall k\in\Real$] 
with $\nu=-1$, 
\eref{e:NLSRHPsoln} reduces to an algebraic system. 
In particular, if $J=1$,
with $k_1= (V + i\,A)/2$ and $C_1= A\,\e^{A\xi+i(\varphi+\pi/2)}$,
one recovers the 
one-soliton solution \eref{e:ivpsoliton}.
In the general reflectionless case with $J>1$, 
\eref{e:NLSRHPsoln} and~\eref{e:reconstq} yield 
[taking into account \eref{e:NLSnorming} and \eref{e:NLSMpmdef}]
the pure multi-soliton solution of the NLS equation as
 \bse
 \label{e:Nsoliton}
 \bea
 q(x,t)= \sum\nolimits_{j=1}^J Z_j\,,
 \\
\noalign{\noindent
where $\@Z= (Z_1,\dots,Z_J)^T$ solves the algebraic system of equations}
 (\_I+\_G)\,\,\@Z= \@c\,,
  \\
\noalign{\noindent
with $\_G=(G_{j,j'})$, $\@c=(c_1,\dots,c_J)^T$,
the superscript $T$ denotes matrix transpose and}
 G_{j,j'}= \e^{2i\theta(x,t,k_j)}C_j
   \sum_{p=1}^J \frac{C_p^*\,\e^{-2i\theta(x,t,k_p^*)}}{(k_j-k_p^*)(k_p^*-k_{j'})}\,,    
 \quad
 c_j= -2i\e^{2i\theta(x,t,k_j)}C_j\,,
 \eea
 \ese
for all $j,j'=1,\dots,J$.

\subsection{Dirichlet and Neumann BCs}
\label{s:DirNeuBCs}

When homogeneous Dirichlet or Neumann BCs are given,
it was shown in~\cite{JMP16p1054}
that the IBVP for the NLS equation can be solved via the IST 
for the whole line using an odd or even extension of the potential. 
In the case of Dirichlet BCs, 
one introduces the odd extension of $\_Q(x)$ in~\eref{e:NLSLP} as
 \be
 \_Q^\ext(x)=\_Q(x)\Theta(x)-\_Q(-x)\Theta(-x) 
 \label{e:oddext}
 \ee
for all $x\in\Real$, where
$\Theta(x)$ is the Heaviside theta function, defined as $\Theta(x)=1$ for
$x > 0$ and $\Theta(x)=0$ for $x<0$. 
Then one defines
$\mu\o{j}(x,t,k)$ through~\eref{e:JostNLS} with $\_Q(x)$ replaced by
$\_Q^\ext(x)$.

It is straightforward to see that the following symmetries hold for
the Jost solutions and, consequently, for the scattering matrix: 
$\forall x\in\Real$ and $\forall k\in\Real$,
 \be
 \mu\o1(x,t,k)=\mu\o2(-x,t,-k)\,,\quad \_A(-k)=\_A^{-1}(k)\,.
 \label{e:symmDir}
 \ee
In particular, $a_{11}(-k)=a_{22}(k)$. 
From the above symmetry and \eref{e:NLSsymmmu},
it then follows that
 \be
 \label{e:Dirsymmscatt}
 a^*(-k^*)=a(k) \quad {\rm and} \quad b(k)=-b(-k)\,.
 \ee
Recall that $J$ is the number of discrete eigenvalues in either 
the upper-half or lower-half complex $k$-plane.
The above relations imply that the discrete spectrum has the following 
properties:
(i)~$J$ is even; (ii)~the discrete eigenvalues appear in quartets, namely
 \be
 \{\pm k_n\,,\,\pm k_n^*\,\}_{n=1}^N\,,
 \label{e:quartets}
 \ee
so that the number of discrete eigenvalues in each quadrant plane is $N= J/2$. 
Moreover, (iii)~the above symmetry in the discrete eigenvalues also induces 
a relation between the corresponding norming constants, as we show next.
Throughout this work, 
we label the discrete eigenvalue symmetric to $k_n$ as
 \be
 k_{n'}=-k_n^*.
 \label{e:eigenvals}
 \ee
We also take $\Re\,k_n\ne0$ for $n=1,\dots,N$ to avoid singular cases.
Using~\eref{e:eigenval} and the symmetries~\eref{e:symmDir}, we have,
for $n=1,\dots,N$,
 \be
 b_nb_{n'}^*=-1\,,\qquad
 C_nC_{n'}^*=1/\.a^2(k_n)\,.
 \label{e:bCDir}
 \ee

The relations between the norming constants take on a particularly
simple form in the reflectionless case.
In particular, for reflectionless potentials and in the case $N=1$
the trace formula yields simply $\.a(k_2)=V/[A(A+iV)]$,
where $k_1= (V+iA)/2$
(see section \ref{s:normingconst} for details). 
%
Writing $C_n= A_n\e^{A_n\xi_n+i(\varphi_n+\pi/2)}$ for $n=1,\dots,2N$,
we then obtain the following explicit relations:
 \be
 \xi_1+\xi_2= \frac1{A}\log \bigg(1+\frac{A^2}{V^2}\bigg)\,, \quad 
 \varphi_2-\varphi_1= 2\arg\big(A+i\,V\big)\,.
 \label{e:Dirnorming}
 \ee

Next, consider the IBVP for~\eref{e:NLS} with Neumann BCs.
In this case one can use the even extension of the potential:
 \be
 \_Q^\ext(x)=\_Q(x)\Theta(x)+\_Q(-x)\Theta(-x)
 \label{e:evenext}
 \ee 
$\forall x\in\Real$. 
As before, one obtains symmetry relations for the eigenfunctions 
and scattering data:
 \be
 \mu\o1(x,t,k)=\sigma_3\mu\o2(-x,t,-k)\sigma_3\,,\quad
 \_A(-k)=\sigma_3\_A^{-1}(k)\sigma_3\,.
 \label{e:symmNeu}
 \ee
For the scattering coefficients we therefore have
 \be a^*(-k^*)=a(k)\,,\quad b(k)=b(-k)\,.
 \label{e:Neusymmscatt}
 \ee
Thus, \eref{e:eigenvals} still applies.
Moreover, one can again obtain relations between the discrete eigenvalues 
and the norming constants: 
 \be
 b_nb_{n'}^*= 1\,,\qquad
 C_nC_{n'}^*= -1/\.a^2(k_n)\,,
 \label{e:bCNeu}
 \ee
for $n=1,\dots,N$.
In particular, for pure soliton solutions and $N=1$ it is
(cf.\ section \ref{s:normingconst}):
 \be
 \label{e:Neunorming}
 \xi_1+\xi_2= \frac1{A}\log \bigg(1+\frac{A^2}{V^2}\bigg)\,, 
 \quad 
 \varphi_2-\varphi_1= 2\arg\big(A+i\,V\big) + \pi\,,
 \ee
where $C_n= A_n\e^{A_n\xi_n+i(\varphi_n+\pi/2)}$ as before.

\subsection{Robin BCs}
\label{s:RobinBCs}

It was shown in \cite{PhysD35p167} that,
even in the case of Robin BCs~\eref{e:RobinBCs},
one can still reduce the IBVP to the solution of an IVP on the whole line.
Motivated by the linear problem (see Appendix), 
one introduces the extension of $\_Q(x)$ as
 \be 
 \_Q^\ext(x,t,k)=\_Q(x,t)\Theta(x)+\_F(k)\_Q(-x,t)\Theta(-x)\,\qquad 
 x\in \Real\,,
 \label{e:Robinext}
 \ee
where $\_F(k)=\diag \big(f(k), f(-k)\big)$, with
 \be 
 f(k)= (2k-i\alpha)/(2k+i\alpha)\,.
 \label{e:fdef}
 \ee
(The factors of 2 in \eref{e:fdef} are chosen for consideration of the 
linear limit, discussed later.)
Since $\_F(k)=\_I$ when $\alpha=0$, and 
$\_F(k)\to-\_I$ as $\alpha\to\infty$,
\eref{e:Robinext} reduces to the odd/even extensions of the potential
in the case of the IBVP with Dirichlet/Neumann BCs, 
\eref{e:oddext} and \eref{e:evenext} respectively.
More generally, 
the extended potential $\_Q^\ext(x,t,k)$ satisfies the symmetries
 \bse
 \label{e:symmQRobin}
 \bea
  \_Q^\ext(-x,t,k)= \_F(-k)\_Q^\ext(x,t,k)\,,
\\
 \Sigma(k)\_F(-k)\_Q^\ext(x,t,k)=-\_Q^\ext(x,t,k)\Sigma(k)\,,
 \eea
 \ese
where $\Sigma(k)= \diag(f(k),-1)$.
 
Let $\mu\o{j}(x,t,k)$ for $j=1,2$ be the Jost solutions
defined via~\eref{e:JostNLS} 
with $\_Q(x,t)$ replaced by $\_Q^\ext(x,t,k)$, as before.
Additional care is necessary for the case of Robin BCs compared to 
Dirichlet and Neumann BCs, because, for all $\alpha\ne0$:
(i)~$\_Q^\ext(x,t,k)$ is discontinuous at $x=0$;
(ii)~for all $x<0$, $\_Q^\ext(x,t,k)$ has simple poles at $k=\pm i\alpha/2$
[since $\_F(k)$ does].
Similar issues arise in the linear limit (cf.\ Appendix).
[There, an extra term must be added to the $t$-part of the Lax pair \eref{e:LSLP} 
to restore its compatibility at $x=0$.  
That is not necessary here, since the Lax pair \eref{e:NLSLP}
is multiplicative, while that of the linear Schr\"odinger equation is additive.]
Note also that, when $r(x,t)= \nu q^*(x,t)$, 
the compatibility of the extended Lax pair \eref{e:NLSLP} 
with $\_Q(x,t)$ replaced by $\_Q^\ext(x,t,k)$ 
implies that $q^\ext(x,t,k)$ solves the NLS equation for all $x\ne0$,
since $f(k)f(-k)= 1$.
As in the linear limit, 
even though $\_Q^\ext(x,t,k)$ is discontinuous at $x=0$, 
the BCs \eref{e:RobinBCs} are satisfied from the left and from the right.

From the symmetries~\eref{e:symmQRobin} one obtains
 \bse
 \label{e:symmRobin}
 \bea
 \mu\o1(x,t,k)=\Sigma(k)\mu\o2(-x,t,-k)\Sigma(-k)\,,
 \label{e:Robinsymmu12}
\\
 \_A(-k)= \Sigma(-k)\_A^{-1}(k)\Sigma(k)\,.
 \label{e:RobinsymmA}
 \eea
 \ese
In particular, 
 \be a(k)=a^*(-k^*),\qquad
 b(k)=f(k)b(-k)\,.
 \label{e:Robinsymmscatt}
 \ee
Equations~\eref{e:symmRobin} and \eref{e:Robinsymmscatt}
generalize the relations found for Dirichlet and Neumann BCs
[namely \eref{e:symmDir}, \eref{e:symmNeu}
and \eref{e:Dirsymmscatt}, \eref{e:Neusymmscatt}, respectively.]

Regarding the analyticity of the eigenfunctions, 
note first that 
$\mu\o2(x,t,k)$ is the same as that of IVP for all $x>0$, 
and therefore enjoys the same properties there.
Using~\eref{e:Robinsymmu12} and definition of $\mu\o1$,
one can obtain the regions of analyticity of
$\mu\o1(x,t,k)$ for $x>0$:
\begin{itemize}
\item
$\mu\o{1,L}$ is analytic for $\Im k<0$, 
except for a simple pole at $k=i\alpha/2$ when $\alpha<0$;
\item
$\mu\o{1,R}$ is analytic for $\Im k>0$,
except for a simple pole at $k=-i\alpha/2$ when $\alpha<0$.
\end{itemize}
[The analyticity properties for $x<0$ can be obtained
using the symmetries~\eref{e:symmQRobin}, \eref{e:symmRobin} and the 
integral equations \eref{e:JostNLS}.
These properties are not necessary for our purposes, however,
since we are only interested in reconstructing the extended potential 
for $x>0$.]
The integral representation of the scattering matrix obtained
from~\eref{e:scatteringNLS} in the limit $x\to\infty$ and
the symmetry~\eref{e:symmQRobin} imply
 \be 
 a(k)= 1 - f(k)\int_0^\infty \mu\o2_{21}(-x,t,-k)r(x,t)\,\d x
   -\int_0^\infty \mu\o2_{21}(x,t,-k)r(x,t)\,\d x\,.
 \ee
Thus, $a(k)$ is analytic for $\Im k>0$, except for a simple pole at 
$k=-i\alpha/2$ when $\alpha<0$.
Thus $\mu\o{1,R}(x,t,k)/a(k)$ in~\eref{e:NLSRHP} has a removable singularity
at $k=-i\alpha/2$. 
By symmetry, the same applies for $\mu\o{1,L}(x,t,k)/a^*(k^*)$ in the 
lower-half plane. 
Therefore, the RHP defined in section~\ref{s:ISTIVP} 
also applies to the case of Robin BCs \eref{e:RobinBCs},
and for all $x>0$, the potential $q(x,t)$ is also reconstructed
in the same way.

Equation \eref{e:Robinsymmscatt} implies that 
the symmetry property \eref{e:eigenvals} of the discrete spectrum
applies in the case of Robin BCs as well.
Moreover, 
similar symmetries as before exist for the norming constants
(again, cf.\ section~\ref{s:normingconst}): 
 \be
 b_nb_{n'}^*= f(k_n)\,,
 \qquad
 C_nC_{n'}^*= - \frac{f(k_n)}{\.a^2(k_n)}\,,
 \label{e:bCrelRobin}
 \ee
for $n=1,\dots,N$.
In particular, 
for pure soliton solutions with $N=1$, the following relations
exist between the norming constants associated to symmetric eigenvalues:
 \bse 
 \label{e:Robinnorming} 
 \bea
 \xi_1+\xi_2= \frac1A\log\bigg(1+\frac{A^2}{V^2}\bigg)
   + \frac1{2A}\log\bigg[\frac{V^2+(A-\alpha)^2}{V^2+(A+\alpha)^2}\bigg]\,,
 \label{e:Robinnorminga} 
 \\
 \varphi_2-\varphi_1= 2\arg\big(A+i\,V\big)
    - \arg\bigg[\frac{V+i(A-\alpha)}{V+i(A+\alpha)}\bigg] + \pi\,.
 \label{e:Robinnormingb}
 \eea \ese
Equations \eref{e:bCrelRobin} and \eref{e:Robinnorming} reduce respectively to 
\eref{e:bCDir} and \eref{e:Dirnorming} as $\alpha\to\infty$ and to 
\eref{e:bCNeu} and \eref{e:Neunorming} as $\alpha\to0$.
In section~\ref{s:solitons} we discuss what \eref{e:eigenvals} and 
\eref{e:bCrelRobin} 
imply about the physical behavior of the solitons in the IBVP.

\section{Relations between discrete eigenvalues and norming constants}
\label{s:normingconst}

We now derive the relations between the discrete eigenvalues and 
the norming constants we presented in section~\ref{s:IBVPNLS}, 
together with their generalization for multi-soliton solutions
and solutions with a nonzero reflection coefficient.
We discuss explicitly the case of Robin BCs
[namely \eref{e:bCrelRobin} and ~\eref{e:Robinnorming}].
The corresponding relations in the case of Dirichlet and Neumann BCs 
[namely, \eref{e:bCDir}, \eref{e:Dirnorming}, \eref{e:bCNeu} 
and~\eref{e:Neunorming}] 
follow from \eref{e:bCrelRobin} and \eref{e:Robinnorming}
by taking the limits $\alpha\to\infty$ and $\alpha\to0$, respectively.

Recall first that the discrete eigenvalues and norming constants 
obey the symmetries \eref{e:NLSsymm0} and \eref{e:quartets}.
From~\eref{e:eigenval} and~\eref{e:Robinsymmu12} 
we have, for all $n=1,\dots,N$,
 \bse
 \label{e:eigenvalk2}
 \bea
 \mu\o{1,R}(x,t,k_{n'})=
   b_{n'}\e^{2i\theta(x,t,k_{n'})}\mu\o{2,L}(x,t,k_{n'})\,,
\\
\noalign{\noindent as well as}
 \mu\o{1,R}(x,t,k_{n'})=
    - \=b_n^{-1}\e^{2i\theta(-x,t,\=k_n)}\Sigma(k_{n'})\mu\o{1,L}(-x,t,\=k_n)\,,
 \eea
 \ese
where $k_{n'}= -k_n^*$ as before.
Also, using~\eref{e:Robinsymmu12} we can write 
 \be
 \mu\o{2,L}(x,t,k_{n'})=
   f(\=k_n)\Sigma^{-1}(\=k_n)\mu\o{1,L}(-x,t,\=k_n)\,.
 \label{e:relation2}
 \ee
Inserting~\eref{e:relation2} into~\eref{e:eigenvalk2} 
leads to
 \be
 \Sigma(k_{n'})\mu\o{1,L}(-x,t,\=k_n)= 
   -\=b_n b_{n'}
    f(\=k_n)\Sigma^{-1}(\=k_n)\mu\o{1,L}(-x,t,\=k_n)\,.
 \ee
Since $f(\=k_n)=1/f(k_{n'})$
and $f(k_{n'})= f^*(k_n)$, we then have $\=b_nb_{n'}=-f^*(k_n)$
for all $n=1,\dots,N$,
which in turn, using \eref{e:NLSsymm0}, yields the first of~\eref{e:bCrelRobin}.

Now recall the definition of the norming constants \eref{e:NLSnorming}
and the symmetry \eref{e:NLSsymm0}.
Using the symmetries of the scattering coefficients \eref{e:NLSsymma}
and \eref{e:Robinsymmscatt},
we obtain $\.a_{11}(\=k_n)=-\.a_{22}(-\=k_n)$ for all $n=1,\dots,2N$.
Hence $\=C_n=b_n^*/\.a_{22}(k_{n'})$, 
and then, noting that $\.a(k_{n'})= - \.a^*(k_n)$, 
we obtain the second of \eref{e:bCrelRobin}.
Note also that, when $C_j= A_j\,\e^{A_j\xi_j+i(\varphi_j+\pi/2)}$,
the second of \eref{e:bCrelRobin} implies
\bse
\label{e:normingrelations}
\bea
\xi_n+\xi_{n'}= \big( \log|f(k_n)| - 2\log|\.a(k_n)| - 2\log A_n \big)/A_n\,,
\\
\varphi_n-\varphi_{n'}= \arg[f(k_n)] - 2\arg[\.a(k_n)] + \pi\,.
\eea
\ese

We now derive \eref{e:Robinnorming} and its generalization to arbitrary
solutions of the IBVP.
It is well known that the analytic scattering coefficients obey
trace formulae.  
Explicitly, for the NLS equation, 
$a(k)=a_{22}(k)$ is given by~\cite{APT2003}:
 \be
 \log a(k)=\sum_{j=1}^J \log\bigg( \frac{k-k_j}{k-k_j^*}\bigg)
   + \frac1{2\pi i}\int_{-\infty}^\infty
     \frac{\log |a(k')|^2}{k'-k}\,\d k'\,,
 \label{e:trace}
 \ee
for all $\Im k>0$.
Using~\eref{e:trace} in \eref{e:normingrelations}
then yields half of the norming constants in terms of the other half. 
In particular, for reflectionless solutions 
the integral in \eref{e:trace} vanishes,
and \eref{e:trace} yields simply
 \be
 \.a(k_j)= \mathop{\prod{}'\kern-0.4em}\limits_{m=1}^J~(k_j-k_m)
   ~\bigg/ \prod_{m=1}^J (k_j-k_m^*)\,
 \label{e:adotgeneralO}
 \ee
for all $j=1,\dots,J$, 
where the prime indicates that the term with $m=j$ is omitted
from the product.
Using the symmetry of the discrete eigenvalues, 
\eref{e:adotgeneralO} becomes
 \bea\fl
 \.a(k_n)= \frac{k_n+k_n^*}{2k_n(k_n-k_n^*)}
   \mathop{\prod{}'\kern-0.4em}\limits_{m=1}^N
     \frac{(k_n-k_m)(k_n+k_m^*)}{(k_n+k_m)(k_n-k_m^*)}
\nonumber\\ 
   = \frac{V_n}{iA_n(V_n+iA_n)}
   \mathop{\prod{}'\kern-0.4em}\limits_{m=1}^N
     \frac{[V_n-V_m+i(A_n-A_m)][V_n+V_m+i(A_n-A_m)]}
        {[V_n+V_m+i(A_n+A_m)][V_n-V_m+i(A_n+A_m)]}\,,
 \label{e:adotgeneral}
 \eea
for all $n=1,\dots,N$.  
One can now substitute \eref{e:adotgeneral} 
into \eref{e:normingrelations}
to obtain the generalization of \eref{e:Robinnorming} as:
\bse
\bea\fl
\xi_n+\xi_{n'}= \frac1{A_n}\log\bigg(1+\frac{A_n^2}{V_n^2}\bigg)
  + \frac1{2A_n}\log\bigg[\frac{V_n^2+(A_n-\alpha)^2}{V_n^2+(A_n+\alpha)^2}\bigg]
\nonumber\\
  - \frac1{A_n}\sum_{m=1}^N{\kern-0.2em}'\log
     \frac{[(V_n-V_m)^2+(A_n-A_m)^2][(V_n+V_m)^2+(A_n-A_m)^2]}
        {[(V_n+V_m)^2+(A_n+A_m)^2][(V_n-V_m)^2+(A_n+A_m)^2]}\,,
\\\fl
\varphi_n-\varphi_{n'}= - 2\arg\big(A_n+i\,V_n\big)
    + \arg\bigg[\frac{V_n+i(A_n-\alpha)}{V_n+i(A_n+\alpha)}\bigg] + \pi
\nonumber\\
  - 2\sum_{m=1}^N{\kern-0.2em}'\arg
       \frac{[V_n-V_m+i(A_n-A_m)][V_n+V_m+i(A_n-A_m)]}
        {[V_n+V_m+i(A_n+A_m)][V_n-V_m+i(A_n+A_m)]}\,.
\eea
\ese
In the simplest case $N=1$ there is no product in \eref{e:adotgeneral},
and the relations between the norming constants become especially simple:
 \[
 C_1C_2^*= (\alpha-A+i\,V)(A-i\,V)^2A^2/[(\alpha+A-i\,V)V^2]\,
 \]
[where $k_1=(V+i\,A)/2$ as before],
which, when inserted in \eref{e:bCrelRobin},
yields~\eref{e:Robinnorming}. 

Note from \eref{e:normingrelations} and \eref{e:adotgeneral} that, 
when $N\ge2$, the norming constants of a given mirror soliton 
are affected by the presence of all the physical solitons.
This is due to the fact that every soliton interaction produces
a position and phase shift, as we discuss next.

\section{Soliton behavior}
\label{s:solitons}

We now discuss the behavior of the soliton solutions of the 
NLS equation on the half line.
We refer to the solitons located to the right of the boundary 
(i.e., at $x>0$) as the \textit{physical} solitons, and 
to the counterparts of the physical solitons to the left
of the boundary as the \textit{mirror} solitons, 
since they can be considered as a reflected image of the physical solitons,
as we will see.
Equation~\eref{e:quartets} obviously implies that the number 
of physical solitons equals that of mirror solitons,
and, with the above notations, this number is $N=J/2$.

\subsection{Soliton reflection}

We first discuss the case $N=1$ for simplicity. 
Solving the algebraic system~\eref{e:Nsoliton} 
one obtains the two-soliton solution of the NLS equation, 
and the solution of the IBVP is then obtained
by choosing the norming constants of the mirror
solitons as explained earlier.


Let $k_1=(V +i\,A)/2$ be the discrete eigenvalue 
of the scattering problem corresponding to the physical soliton.
Recall from~\eref{e:ivpsoliton} that the real part and imaginary part 
of the discrete eigenvalue determine respectively the velocity and 
the amplitude of the soliton. 
By the symmetry~\eref{e:eigenvals} of the discrete spectrum, 
we know that $k_2=-k_1^*= (-V+i\,A)/2$ is also an eigenvalue,
corresponding to the mirror soliton.
Hence, the mirror soliton has the same amplitude as, and opposite velocity to,
the physical soliton.

\begin{figure}[t!]
\rightline{\raise1.2ex\hbox{%
\includegraphics[width=0.475\textwidth]{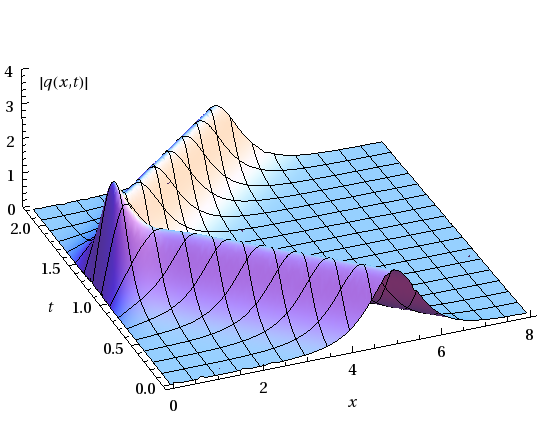}}\kern1em
\includegraphics[width=0.355\textwidth]{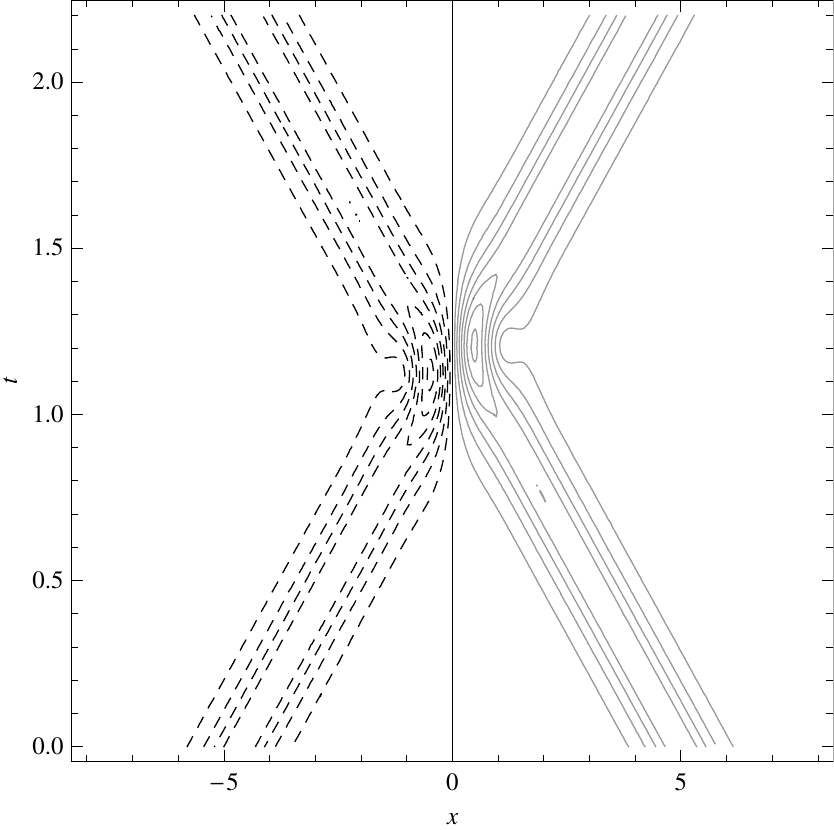}}
\caption{Soliton reflection at the boundary in the case of Dirichlet BCs,
with $A=2$, $V=-2$, $\xi=5$, and $\varphi=0$.
Left: three-dimensional (3D) plot of $|q(x,t)|$.
Right: contour plot showing the mirror soliton (dashed) 
to the left of the boundary.} 
\label{f:nlsD}
\vskip3\medskipamount
\rightline{\raise1.2ex\hbox{%
\includegraphics[width=0.475\textwidth]{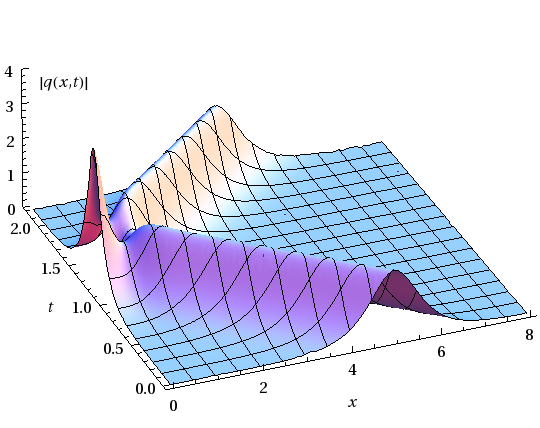}}\kern1em
\includegraphics[width=0.355\textwidth]{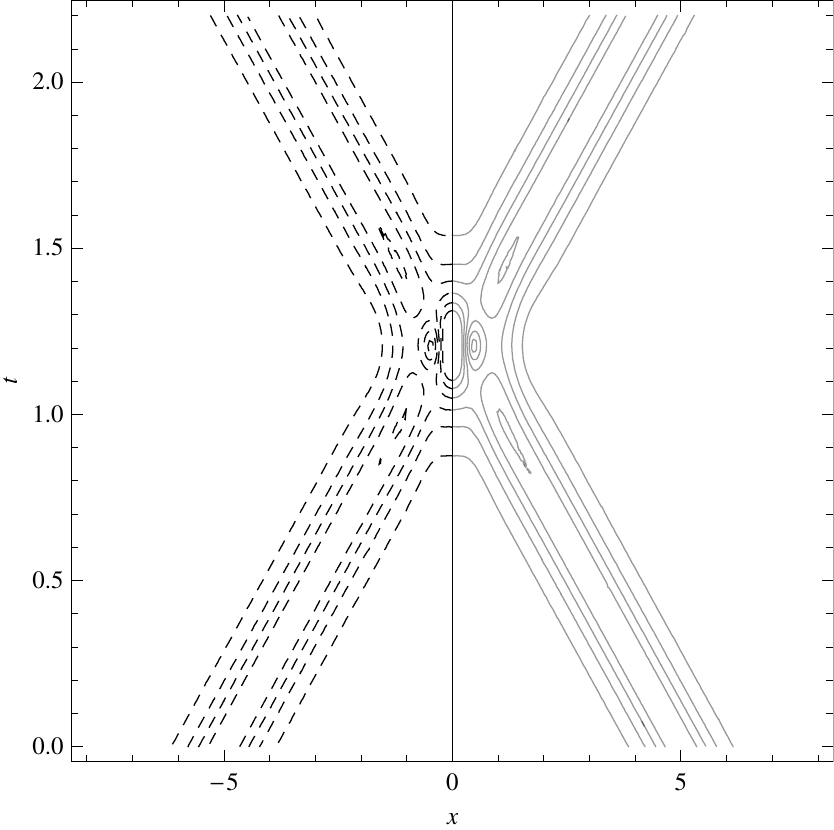}}
\caption{Soliton reflection at the boundary in the case of Neumann BCs,
with $A=2$, $V=-2$, $\xi=5$, and $\varphi=0$.
Left: 3D plot.
Right: contour plot showing the mirror soliton.}
\label{f:nlsN}
\vskip3\medskipamount
\rightline{\raise1.2ex\hbox{%
\includegraphics[width=0.475\textwidth]{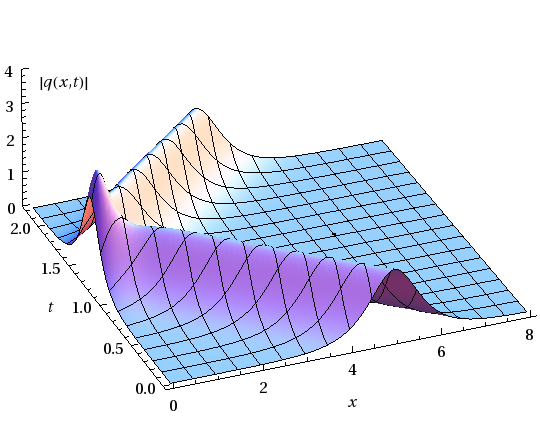}}\kern1em
\includegraphics[width=0.355\textwidth]{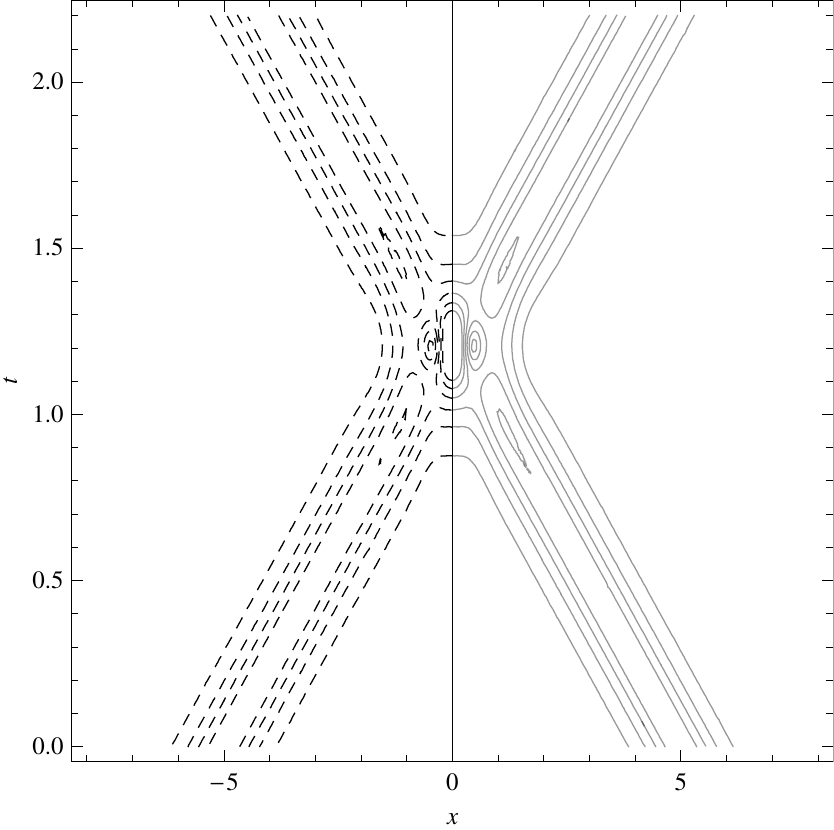}}
\caption{Soliton reflection at the boundary in the case of Robin BCs
with $\alpha=3$, and with
$A=2$, $V=-2$, $\xi=5$, and $\varphi=0$.
Left: 3D plot.
Right: contour plot showing the mirror soliton.}
\label{f:nlsR}
\kern-2\medskipamount
\end{figure}

Figure~\ref{f:nlsD} shows (left) the soliton reflection at the boundary
in the case of Dirichlet BCs, together with a contour plot (right)
that includes the mirror soliton (dashed lines).
The above results imply that 
the soliton reflection at $x=0$ is simply a particular case of an 
elastic two-soliton interaction of the NLS equation, in which
the norming constant of the mirror soliton is chosen so as to
make the whole solution zero at the origin.
As usual in a soliton interaction, the solitons re-emerge intact 
after the collision, 
except that in our case the roles of physical and mirror soliton 
are now swapped.
A similar scenario occurs in the case of Neumann and Robin BCs, 
as shown respectively in Figs.~\ref{f:nlsN} and \ref{f:nlsR}, 
except that the norming constant of the mirror soliton in each case 
is such that the appropriate BCs are satisfied.

\begin{figure}[t!]
\kern-\medskipamount
\rightline{\raise1.2ex\hbox{%
\includegraphics[width=0.475\textwidth]{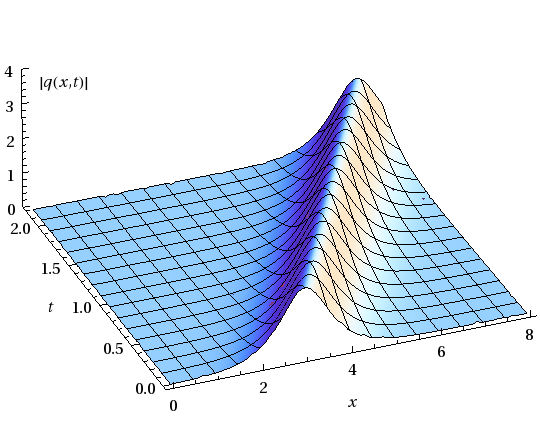}}\kern1em
\includegraphics[width=0.355\textwidth]{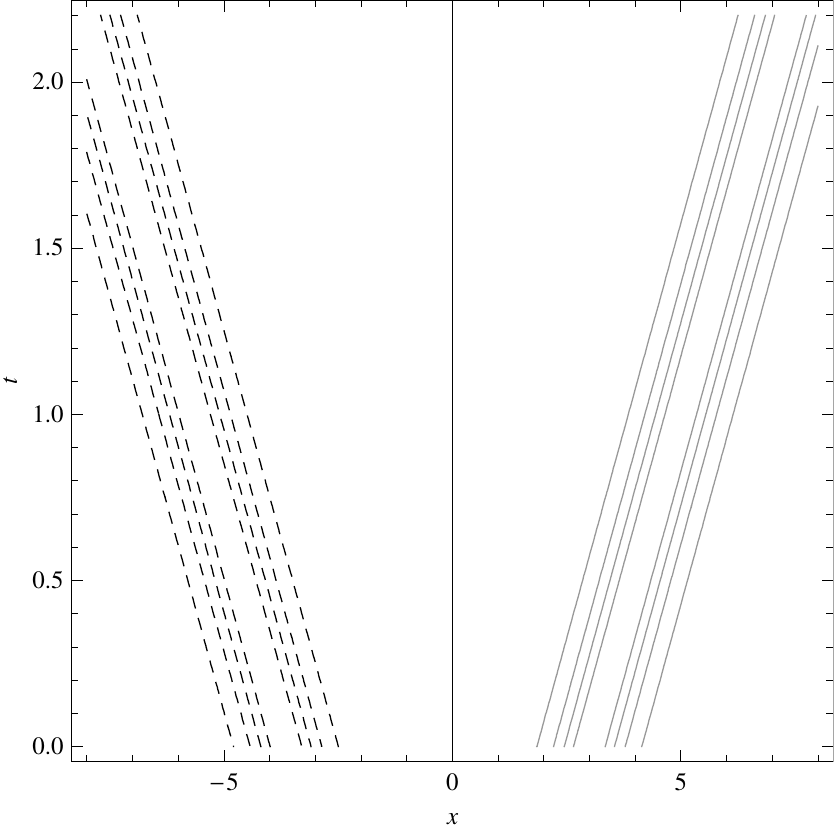}}
\caption{Reflection at the boundary in the case of Robin BCs
and positive velocity:
$\alpha=3$, $A=2$, $V=1$, $\xi=3$, and $\varphi=0$.
Left: 3D plot.
Right: contour plot showing the mirror soliton.
In this case the solution with Dirichlet and Neumann BCs 
is visually undistinguishable from the above.}
\label{f:nlsR2}
\kern2\medskipamount
\rightline{\raise1.4ex\hbox{%
\includegraphics[width=0.485\textwidth]{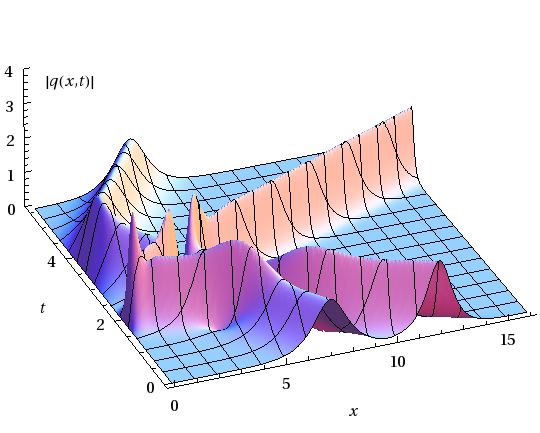}}\kern1em
\includegraphics[width=0.335\textwidth]{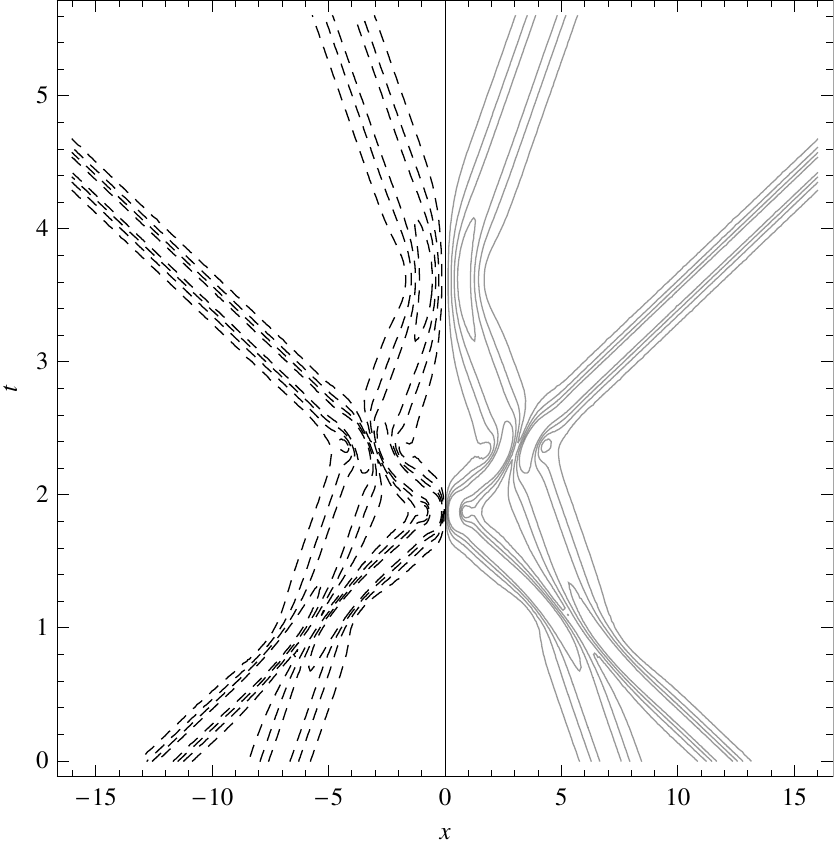}}
\caption{Reflection at the boundary of two physical solitons:
$A_1=2$, $A_2=3/2$, $V_1=-3$, $V_2=-1$, $\xi_1=12$, $\xi_2=8$,
and $\varphi_1=\varphi_2=0$.
Left: 3D plot.  Right: contour plot.}
\label{f:2soliton}
\kern2\medskipamount
\rightline{\raise1.4ex\hbox{%
\includegraphics[width=0.485\textwidth]{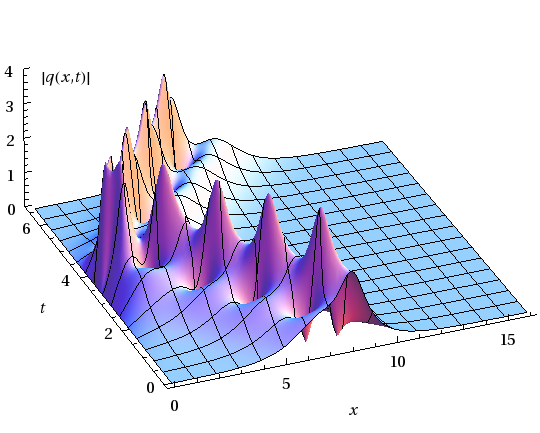}}\kern1em
\includegraphics[width=0.335\textwidth]{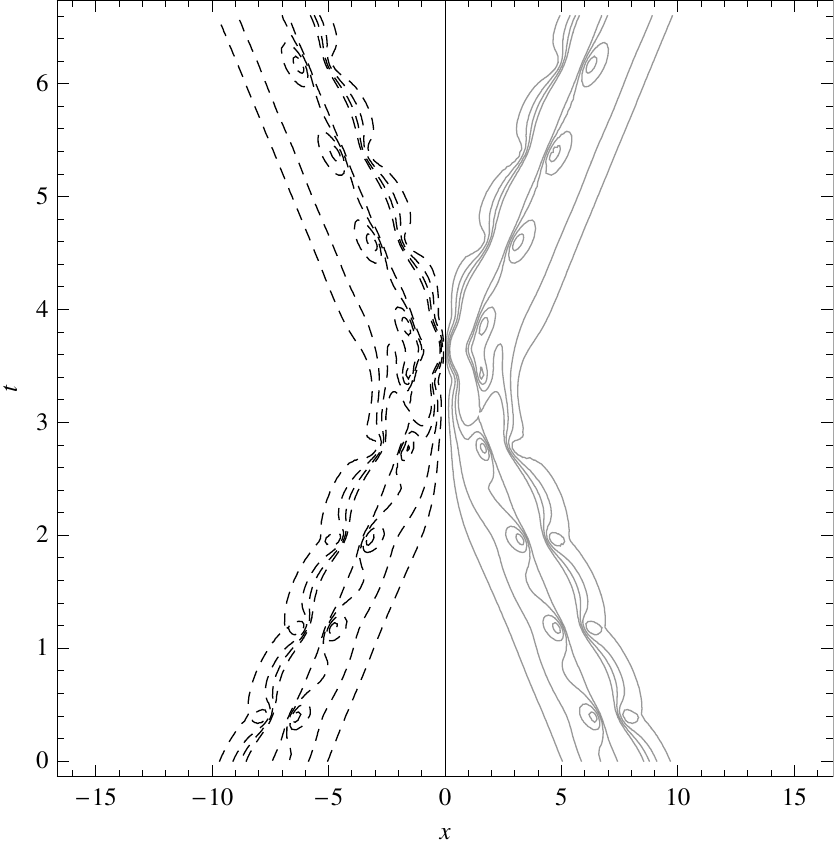}}
\caption{Reflection at the boundary of two physical solitons:
$A_1=1$, $A_2=3$, $V_1=V_2=-1$, $\xi_1=\xi_2=10$,
and $\varphi_1=\varphi_2=0$.
Left: 3D plot.  Right: contour plot.}
\label{f:boundstate}
\kern-2\medskipamount
\end{figure}

Note that the symmetry of the discrete spectrum and
the relations between eigenvalues and norming constants 
apply independently of whether the physical soliton has a positive 
or negative velocity.
In other words, they apply whether the discrete eigenvalue associated to
the physical soliton is located in the first or second quadrant of the 
complex $k$-plane [recall $V_n= 2\Re k_n$].
Of course, if the physical soliton has a positive velocity, 
no soliton reflection occurs for $t>0$, and the solution is
exponentially small at the origin for all $t>0$.
Nonetheless, a mirror soliton is still needed to satisfy the BCs
at the origin, as shown in Fig.~\ref{f:nlsR2}.

The location of the mirror soliton is the same for Dirichlet and Neumann BCs, 
and in those two cases only the phase difference between the solitons contributes 
to satisfying the BCs [cf.\ \eref{e:Dirnorming} and \eref{e:Neunorming}].
The same is not true, however, for the more general BCs \eref{e:RobinBCs}
with $\alpha\ne0$ [cf.\ \eref{e:Robinnorming}].

Finally, figure~\ref{f:2soliton} displays the reflection of
two physical solitons in the case of Dirichlet BCs,
demonstrating that our results are not limited to the case $N=1$.
Similarly, figure~\ref{f:boundstate} displays the reflection of a two-soliton
bound state, also in the case of Dirichlet BCs.
As before, when $N>1$ one first solves the algebraic system~\eref{e:Nsoliton}
to obtain the $2N$-soliton solution of the NLS equation.
The solution of the IBVP is then obtained by appropriately choosing 
the norming constants of the mirror solitons.
The generalizations of \eref{e:Robinnorming} to obtain the
norming constants for solutions with $N\ge2$ 
and for solutions with non-zero reflection coefficients are
described in section~\ref{s:normingconst}.


It is convenient to label the discrete eigenvalues such that
$\Re k_1\le \Re k_2\le \cdots \le \Re k_{2N}$.
With this convention, 
as $t\to-\infty$ all physical solitons correspond to the 
$N$ discrete eigenvalues in the second quadrant of the complex $k$-plane
(i.e., those with $\Re k_n<0$).
Then, as each soliton is reflected in succession,
the corresponding discrete eigenvalue associated with the physical
soliton switches role with its symmetric conuterpart,
until, as $t\to\infty$, all physical solitons correspond to the 
$N$ discrete eigenvalues in the first quadrant (i.e., with $\Re k_n>0$).
Consequently, as $t\to-\infty$ the discrete eigenvalues associated with 
the physical solitons as $k_1,\dots,k_N$, and the corresponding 
mirror solitons are given respectively by $k_{2N},\dots,k_{N+1}$.
That is, $n'= 2N-n+1$ for all $n=1,\dots,N$.
Conversely, as $t\to\infty$ the eigenvalues associated with the 
physical solitons are $k_{N+1},\dots,k_{2N}$, and the corresponding 
mirror solitons are given respectively by $k_N,\dots,k_1$.
That is, $n'= 2N-n+1$ for all $n=N+1,\dots,2N$.

\subsection{Reflection-induced shift}

Recall that the soliton reflection at the boundary is effectively
the interaction between the physical soliton and its mirror image.
Since any soliton interaction results in a position shift,
it follows that the soliton reflection at the boundary also produces 
such a shift.
As we show next, however, 
a second contribution also exists to the total reflection-induced shift.
In fact, depending on the soliton parameters, 
this second contribution can even make the total shift zero 
as if the whole process were purely linear. 

It is well-known that, as $t\to\infty$,
a multi-soliton solution becomes asymptotically a linear superposition of 
one-soliton solutions \cite{APT2003,JETP34p62}.
That is, 
 \be
q(x,t) \sim \sum_{j=1}^J q_j^\pm(x,t)\,
 \ee
as $t\to\pm\infty$, 
where $q_j^\pm(x,t)$ is of the form \eref{e:ivpsoliton}, but with the
soliton parameters $A,V,\xi,\varphi$ replaced by $A_j,V_j,\xi_j^\pm,\varphi_j^\pm$
for $j=1,\dots,J$.
The fact that $\xi_j^\pm$ and $\varphi_j^\pm$
do not coincide with each other is the manifestation of the interaction-induced shift.
%
%
Define as usual the interaction-induced position shift of the 
$j$-th soliton as $\delta \xi_j= \xi_j^+-\xi_j^-$, and 
label the discrete eigenvalues so that $V_1<V_2<\cdots<V_J$.
Without repeating the calculations \cite{APT2003,JETP34p62}, 
we quote the relevant results:
for all $j=1,\dots,J$ it is
\bea\fl
A_j\xi_j = A_j\xi_j^+ - \log|\.a(k_j)|
      - \sum_{m=1}^J\!\!{ }'\,\,\sigma_{m,j}\log|a_m(k_j)|
    = A_j\xi_j^- - \log|\.a(k_j)|
      + \!\sum_{m=1}^J \!\!{ }'\,\,\sigma_{m,j}\log|a_m(k_j)|\,,
\nonumber\\[-1ex]
\label{e:asymptotics}
\eea
where 
$a_j(k)= (k-k_j)/(k-k_j^*)$ is the transmission coefficient 
for a one-soliton solution [cf.\ \eref{e:trace}],
$\sigma_{m,j}= -1$ for $m=1,\dots,j-1$ and 
$\sigma_{m,j}= 1$ for $m=j+1,\dots,J$,
and as before the sum is taken over all $m\ne j$.
Comparing the asymptotic results as $t\pm\infty$, we then obtain
the position shift for multi-soliton solutions of the NLS equation as 
 \be
 \delta\xi_j=
   \frac2{A_j}\sum_{m=1}^J \!\!{ }'\,\,
      \sigma_{m,j}\log\bigg|\frac{k_j-k_m}{k_j-k_m^*}\bigg|\,.
 \label{e:totalinteractionshift}
 \ee
Note that  the position shifts are pairwise additive.
That is, 
$\delta\xi_j= \sum\nolimits_{m=1}^J \delta \xi_{j,m}$,
where $\delta\xi_{j,m}$ 
is the position shift of the $j$-th soliton arising from 
its interaction with the $m$-th soliton.
In particular, if $k_n=(V_n+iA_n)/2$ with $V_n<0$
is the eigenvalue associated to the physical soliton at $t=0$, 
and $k_{n'}=-k_n^*$ is the mirror eigenvalue,
\eref{e:totalinteractionshift} yields the position shift resulting from 
the interaction of the physical soliton with its mirror: 
 \bea
 \delta\xi_{n,n'} = 
   - \frac1{A_n}\log\Bigg( 1 + \frac{A_n^2}{V_n^2}\Bigg) =
   - \delta\xi_{n',n}\,.
 \label{e:interactionshift}
 \eea

\begin{figure}[t!]
\rightline{\lower1ex\hbox{\includegraphics[width=0.375\textwidth]{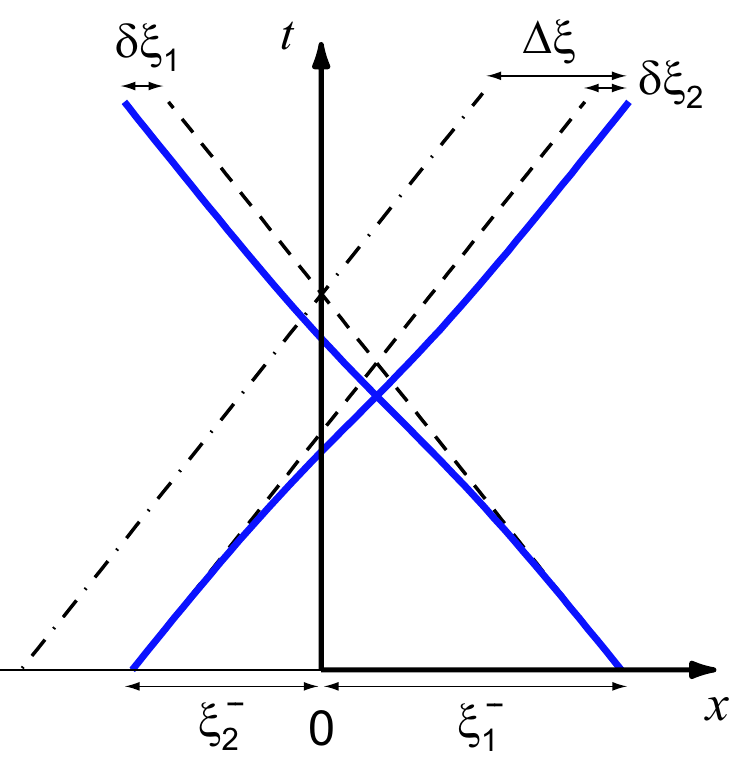}}\qquad
\raise1ex\hbox{\includegraphics[width=0.35\textwidth]{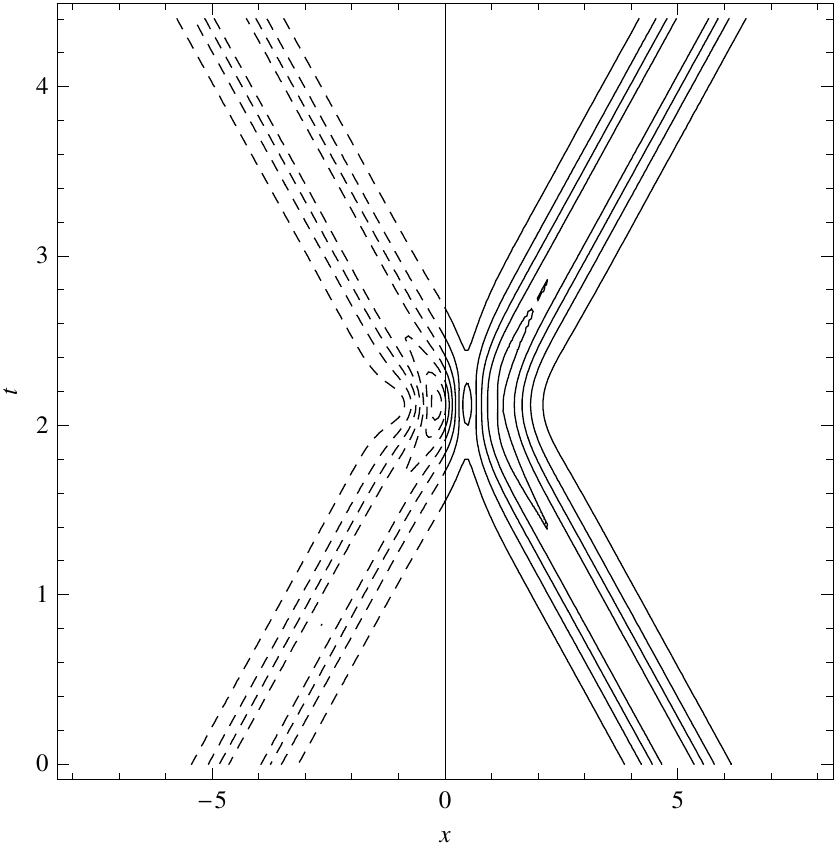}}}
\kern-\smallskipamount
\caption{Left: schematic diagram of the reflection-induced shift $\Delta\xi$.
The displacement from the origin of the interaction center has been 
exaggerated for illustration purposes.
Right: a soliton reflection with a large, boundary-induced shift:
$\alpha=-2$, $A=2$, $V=-1$, $\xi=5$ and $\varphi=0$.}
\label{f:timingshift}
\end{figure}

Since the physical and mirror soliton interchange roles in the reflection, 
however,
the total reflection-induced shift is not simply due to the soliton 
interaction.
More precisely, 
define the reflection-induced position shift $\Delta\xi$ as the 
displacement of the soliton center from where it would be 
had it followed a purely piecewise linear path
[cf.\ Fig.~\ref{f:timingshift}].
A simple calculation 
shows that such a position shift is given by
 \be
 \Delta \xi_n = \xi_n^-+\xi_{n'}^- - \delta\xi_n \,,
 \label{e:reflectionshift}
 \ee
where as before $\delta\xi_n$ the interaction-induced shift,
obtained from \eref{e:totalinteractionshift},
and the index $n'$ labels the mirror soliton of the $n$-th soliton.
As is evident from~\eref{e:asymptotics},
the asymptotic soliton parameters 
$\xi_j^\pm$ and $\varphi_j^\pm$ do not coincide with the constants
$\xi_j$ and $\varphi_j$ appearing in the exact $J$-soliton solution 
\eref{e:Nsoliton}.
In particular, using the symmetry of the discrete eigenvalues,
after some tedious but straightforward algebra 
\eref{e:asymptotics} yields simply
\bea\fl
\xi_n^\pm + \xi_{n'}^\pm = \xi_n + \xi_{n'} + 2\log|\.a(k_n)|
  = \log|f(k_n)| - 2 \log A_n
   = \frac1{A_n}
       \log\bigg[ \frac1{A_n^2}
	 \bigg(\frac{V_n^2+(A_n-\alpha)^2}{V_n^2+(A_n+\alpha)^2}\bigg)^{1/2}
   \bigg]\,,
\nonumber\\[-1ex]
\label{e:xipm}
\eea
where \eref{e:normingrelations} was used.
The reflection-induced shift $\Delta\xi$ is then obtained inserting
\eref{e:interactionshift} and \eref{e:xipm} 
into \eref{e:reflectionshift}.
In particular, for $N=1$, it is:
\be
\Delta\xi= \frac1A\bigg[ \log\bigg(\frac1{A^2} + \frac1{V^2}\bigg)
  + \frac12\log\bigg(\frac{V^2+(A-\alpha)^2}{V^2+(A+\alpha)^2}\bigg)\bigg]\,.
\label{e:N=1reflectionshift}
\ee

Importantly, the reflection-induced shift 
depends on the BCs.
Indeed, such a dependence is evident in 
Figs.~\ref{f:nlsD}, \ref{f:nlsN}, \ref{f:nlsR} and \ref{f:timingshift}.
It should also be clear than $\Delta\xi$ can be either positive or negative 
depending on the soliton parameters and BCs.
Note also that, if the physical soliton has a positive velocity
(i.e., if $V>0$), 
the reflection at the boundary and the corresponding position shift
obviously occur at $t<0$.

The above results are easily generalized to the case $N\ge2$.
Indeed, \eref{e:xipm} holds for all~$N$.
Also, taking $n\le N$ (i.e., $V_n<0$), using the symmetry of the eigenvalues,
\eref{e:asymptotics} yields 
\bea\fl
\delta\xi_n=  - \frac1{A_n}\log\Bigg( 1 + \frac{A_n^2}{V_n^2}\Bigg)
  + \frac1{A_n}\sum_{m=1}^N{\!}'\,\log\bigg[
     \frac{(V_n-V_m)^2+(A_n-A_m)^2}{(V_n-V_m)^2+(A_n+A_m)^2}\bigg]
\nonumber\\\kern8em{ }
  + \frac1{A_n}\sum_{m=1}^N{\!}'\,\sigma_{m,n}\log\bigg[
     \frac{(V_n+V_m)^2+(A_n-A_m)^2}{(V_n+V_m)^2+(A_n+A_m)^2}\bigg]\,.
\label{e:symminteractionshift}
\eea
One can then combine \eref{e:xipm} and \eref{e:symminteractionshift}
into \eref{e:reflectionshift}
to obtain the generalization of \eref{e:N=1reflectionshift}.
\eject

Note that, when $N\ge2$, 
the interaction shifts $\delta\xi_n$ compare the soliton positions
before and after all interactions have occurred.
Similarly, 
the constants $\xi_n^\pm$ and $\varphi_n^\pm$ apply for times respectively 
before and after all soliton interactions (and thus reflections) 
have occurred.
So \eref{e:reflectionshift} applies in that limit.
But equations \eref{e:bCrelRobin} allow one to obtain the 
$N$-soliton solution of the IBVP given the $N$ discrete eigenvalues
and norming constants associated with the physical solitons at any value of $t$,
whichever these may be, 
and independently of how many reflections may already have occurred.

\section{Discussion}
\label{s:discussion}

Since many physical situations naturally give rise to IBVPs for the 
NLS equation,
we expect that our characterization of the soliton solutions of the IBVP 
will have a broad range of applicability.
On the technical side we note that the approach we used for the IBVP
--- namely, extension of the potential and use of the IST for the IVP
\cite{JMP16p1054,PhysD35p167} ---
is fundamentally different from the new method for IBVPs presented 
in~\cite{PRSLA453p1411}, which is based on the simultaneous spectral
analysis of both parts of the Lax pair.
We also note that the symmetries of the scattering coefficients 
had been derived in~\cite{JMP16p1054} for Dirichlet and Neumann BCs 
and in~\cite{PhysD35p167} for Robin BCs (see also~\cite{JPA24p2507,IP7p435}).
None of those works, however, discussed the symmetries of the 
discrete spectrum, norming constants and the corresponding implications 
on the soliton behavior.

We should also emphasize that the symmetries of the scattering data and 
the discrete spectrum only apply to linearizable BCs.
In fact, a trivial counterexample in the case of other BCs is given by
\eref{e:ivpsoliton}, which provides a perfectly valid solution to the IBVP 
for the NLS equation on the half line with the 
non-homogeneous Dirichlet BC $q(0,t)=q_{\rm s}(0,t)$.

Finally, we note that, similarly to the traditional method of images, 
one could in principle \textit{assume} that the symmetry 
\eref{e:quartets} of the discrete eigenvalues holds, 
require that the solution satisfies the given BCs 
and then obtain the relation between the norming constants
by solving a system of algebraic equations.
We did not do so here, however.  
On one hand, the resulting equations are transcendental, 
and finding their solution without any a priori knowledge 
is highly nontrivial, which would make the method impractical.
Moreover, exploiting the symmetry of the scattering coefficients
enabled us to \textit{prove} that \textit{all} solutions of the IBVP 
possess the symmetry \eref{e:quartets} and to find the relations 
between norming constants, regardless of the number of
physical solitons and of whether the reflection coefficient is zero. 

The present results open up a number of interesting questions. 
An obvious one is what happens for other integrable NLEEs.
We expect that similar results will apply to equations 
such as sine-Gordon equation \cite{JPA37L471} and the 
Ablowitz-Ladik lattice \cite{IP24p065011}.
Less clear, however, is what happens for the Korteweg-deVries equation, 
since in this case all solitons travel to the right (or to the left
depending on the sign of the nonlinear term), and no obvious symmetry 
exists.
It might be that the nonlinear method of images only applies
as presented to equations that admit some reflection symmetry, or such that
the linear limit can be solved with Fourier methods \cite{PRSLA456p805}.
(Note that even the traditional method of images has similar restrictions.) 

Another nontrivial question is what happens 
on finite domains.  
Numerical simulations show that in this case the solitons experience
an infinite number of reflections at each boundary.
After two reflections, however, any soliton recovers its original
velocity.
The initial soliton and its twice-reflected copy must thus be associated 
to the same discrete eigenvalue. 
Moreover, since the soliton experiences an infinite number of reflections,
the extension of the solution to the whole line would require an 
infinite number of mirror copies.
Finally, such a solution would not possess decaying or even constant BCs 
as $x\to\pm\infty$.  
Therefore, its characterization seems to be outside the current capabilities 
of IST-based methods.
\eject

\section*{Acknowledgements}

We thank M J Ablowitz, A S Fokas and W L Kath for many interesting discussions.
This work was partially supported by the National Science Foundation
under grant DMS-0506101.

\def\thesubsection{\Alph{section}.\arabic{subsection}}
\def\numparts{\refstepcounter{equation}%
     \setcounter{eqnval}{\value{equation}}%
     \setcounter{equation}{0}%
     \def\theequation{\Alph{section}.\arabic{eqnval}{\it\alph{equation}}}}
\def\endnumparts{\def\theequation{\Alph{section}.\arabic{equation}}%
     \setcounter{equation}{\value{eqnval}}}
\def\theequation{\Alph{section}.\arabic{equation}}
\def\thesection{Appendix~\Alph{section}}
\setcounter{section}1
\section*{Appendix: IBVP for the linear Schr\"odinger equation with Robin BCs}
\label{s:LIBVPRobinBC}

For comparison purposes, 
here we solve the IBVP for the linear Schr\"odinger (LS) equation 
 \be 
 iq_t+q_{xx}=0\, 
 \label{e:LS}
 \ee
for $x>0$, with the homogeneous Robin BCs~\eref{e:RobinBCs} at the origin,
using the IST for the IVP and an extension 
of the potential to the negative real $x$-axis.
(See \cite{IP24p065011,PRSLA453p1411} for an alternative method.)

Equation~\eref{e:LS} is the compatibility condition 
of the scalar Lax pair~\cite{IP24p065011}
 \be 
 \phi_x - ik\phi = q\,, \qquad 
 \phi_t + ik^2\phi = iq_x-k q\,. 
 \label{e:LSLP}
 \ee
When \eref{e:LS} is posed on $-\infty<x<\infty$, 
the Jost solutions of~\eref{e:LSLP} are
 \bea
 \phi^+(x,t,k)=\int_{-\infty}^x \e^{ik(x-x')}q(x',t)\,\d x'\,,
 \quad
 \phi^-(x,t,k)= -\int_x^\infty \e^{ik(x-x')}q(x',t)\,\d x'\,.
 \nonumber\\[-2ex]
 \label{e:LSJost}
 \eea
It is easily shown that $\phi^\pm(x,t,k)$ can be analytically 
continued on $\Im k\gl0$, and $\phi^\pm(x,t,k)=O(1/k)$ as $k\to\infty$.
Moreover, on $\Im k=0$ one has the following jump condition:
 \be 
 \phi^+(x,t,k)-\phi^-(x,t,k)= \e^{ikx}\^q(k,t)\,,
 \label{e:RHPLS}
 \ee 
with $\^q(k,t)$ given by the first of \eref{e:FTpair} below.
Equation~\eref{e:RHPLS}
defines a scalar RHP that is solved with standard Cauchy projectors.
Moreover, the jump data satisfies 
 \be
 \^q(k,t)= \e^{-ik^2t}\^q(k,0)
 \label{e:evolscatt}
 \ee 
and the solution of the RHP leads to the direct and inverse 
Fourier transform pair:
 \be 
 \^q(k,t)= \int_{-\infty}^\infty \e^{-ikx}q(x,t)\,\d x\,,
 \qquad
 q(x,t)= \frac1{2\pi}\int_{-\infty}^\infty \e^{ikx}\^q(k,t)\,\d k\,.
 \label{e:FTpair}
 \ee

Now consider the IBVP for \eref{e:LS} on $x>0$ with BCs \eref{e:RobinBCs}.
It is trivial to see that, if $q(x,t)$ is given by the second of 
\eref{e:FTpair} for $x>0$,
the BCs \eref{e:RobinBCs} are satisfied provided that
 \bea
 \^q(k,t)=f_o(k)\,\^q(-k,t)\,\,,
 \label{e:relationqBC}
 \\[-2ex]
\noalign{\noindent where}
 f_o(k)= (k-i\alpha)/(k+i\alpha)\,.
 \eea
This motivates the following extension of $q(x,t)$:
 \be 
 q^\ext(x,t,k)=q(x,t)\Theta(x)+f_o(k)q(-x,t)\Theta(-x)\,,\quad 
 \label{e:qextdef}
 \ee
for all  $x\in \Real$.
The corresponding Fourier transform $\^q^\ext(k,t)$ 
[defined by the first of \eref{e:FTpair} 
with $q(x,t)$ replaced by $q^\ext(x,t)$]
is related to the the one-sided Fourier transform $\^q_o(k,t)$
 \be
 \^q_o(k,t)= \int_0^\infty \e^{-ikx}q(x,t)\,\d x\,,
   \ee
by $\^q^\ext(k,t)= \^q_o(k,t) + f_o(k)\^q_o(-k,t)$.
It is then easily verified that $\^q^\ext(k,t)$
satisfies~\eref{e:relationqBC} [note $f_o(k)f_o(-k)=1$]. 
We then consider an extended Lax pair obtained by replacing 
$q(x,t)$ with $q^\ext(x,t,k)$ in~\eref{e:LSLP}.
Note however that $q^\ext(x,t,k)$ is in general discontinuous at $x=0$, 
and in a distributional sense it solves
 \bea 
 iq^\ext + q_{xx}^\ext = [f_o(k)+1]\,q_x(0,t)\delta(x)
   -[f_o(k)-1]\,q(0,t)\delta'(x)\,, 
 \nonumber
 \eea
where $\delta(x)$ is the Dirac delta.
Thus the Lax pair~\eref{e:LSLP} with $q(x,t)$ replaced by 
the extended potential is not compatible at $x=0$. 
The compatibility is restored if the $t$-part of~\eref{e:LSLP}
is replaced by
 \be 
 \phi_t+ik^2\phi=iq_x^\ext-kq^\ext +i\,\big[f_o(k)-1\big]q(0,t)\delta(x)\,.
 \label{e:correctedtLP}
 \ee
One now defines $\phi^\pm(x,t,k)$ as the Jost solutions of this new Lax pair.
Since the $x$-part is formally the same as before,
they are simply given by~\eref{e:LSJost} 
with $q(x,t)$ replaced by $q^\ext(x,t,k)$.
For $x>0$, $\phi^-(x,t,k)$ is analytic on $\Im k <0$. 
Also, for $x>0$, $\phi^+(x,t,k)$ is analytic on $\Im k >0$ 
when $\Re\alpha\ge0$ and is meromorphic on $\Im k>0$ 
with a simple pole at $k=-i\alpha$ when $\Re\alpha<0$.
[When $\Re\alpha>0$, the pole at $k=-i\alpha$ is in the lower-half plane.] 
The jump condition \eref{e:RHPLS} still holds 
with $\^q(k,0)$ replaced by $\^q^\ext(k,0)$.
Moreover, 
using~\eref{e:correctedtLP} with $q(x,t)\to 0$ as $x\to\infty$, 
we obtain that the scattering data $\^q^\ext(k,t)$
still evolves according to \eref{e:evolscatt}.
After subtracting the pole contribution, one solves the RHP~\eref{e:RHPLS}
using Cauchy projectors, obtaining:
 \bea 
 \phi(x,t,k)= \frac1{2\pi i}\,
 \int_{-\infty}^\infty
  \e^{i(k'x-k'^2t)}\,\frac{\^q^\ext(k',0)}{k'-k}\,\d k'
    + \frac1{k+i\alpha}\,\nu'_\alpha \Res_{k=-i\alpha}\big[\phi^+\big]\,,
  \nonumber
 \eea
where 
$\nu'_\alpha=0$ for $\Re\,\alpha>0$, 
$\nu'_\alpha =1/2$ for $\Re\,\alpha=0$, 
and $\nu'_\alpha=1$ for $\Re\,\alpha<0$.
Note that 
$\Res\nolimits_{k=-i\alpha}[\phi^+]=
  -2i\alpha\e^{\alpha x+i\alpha^2 t}\^q(i\alpha,0)$. 
Substituting the above into~\eref{e:LSLP} we have, for all $x>0$,
 \be 
 q(x,t)= \frac1{2\pi}\,\int_{-\infty}^\infty\e^{i(kx-k^2t)}\^q^\ext(k,0)\,\d k
   - 2\nu'_\alpha\alpha \e^{\alpha x+i\alpha^2 t}\^q_o(i\alpha,0)\,,  
 \label{e:LSRobinBCsoln3}
 \ee
Note that, even though $q^\ext(x,t,k)$ is discontinuous at $x=0$, 
the BCs \eref{e:RobinBCs} are satisfied from the left and from the right.

One can compare the above solution to the linear limit 
of the solution of the IBVP for the NLS equation.
To do so, let $\_Q^\ext(x,t,k)=\_Q^\ext(x,t,k;\epsilon)=O(\epsilon)$
in~\eref{e:NLSLP}. 
As $\epsilon\to 0$, it is:
 \bse
 \bea
 a(k)=1+O(\epsilon^2)\,,\quad 
 b(k)=\int_{-\infty}^\infty\e^{2ikx}q^\ext(x,t) \,\d x + O(\epsilon^2)\,,
 \\[-1ex]
 \mu_{12}\o1(x,t,k)=\int_0^x\e^{2ik(x-x')}q(x',t)\,\d x' 
   + f(k)\int_0^\infty\e^{2ik(x+x')}q(x',t)\,\d x'+O(\epsilon^2)\,.
 \nonumber\\[-2ex]
 \eea
 \ese
Since $\mu\o{1,R}(x,t,k)$ has a pole at $k=-i\alpha/2$ when $\alpha<0$
but $a(k)=1$ to leading order,
the solution of the RHP~\eref{e:NLSRHP} acquires an 
additional term generated by the residue of $\mu\o{1,R}(x,t,k)/a(k)$.
Taking the pole contribution into account, and performing 
the change of variable, $2k\to k'$, one then obtains, 
to $O(\epsilon)$, \eref{e:LSRobinBCsoln3}.
That is, in the linear limit, the solution of the IBVP for the NLS equation 
coincides with solution of the IBVP for the linear Schr\"odinger equation.

\catcode`\@ 11
\def\journal#1&#2,#3 (#4){\begingroup \let\journal=\d@mmyjournal {\frenchspacing\sl #1\/\unskip\,}
{\bf\ignorespaces #2}\rm, #3 (#4)\endgroup}
\def\d@mmyjournal{\errmessage{Reference foul up: nested \journal macros}}
\def\title#1{{``#1''}}
\def\@biblabel#1{#1.}

\end{document}